\documentclass[twocolumn]{aastex62}

\newcommand{\msunh}{\>h^{-1}\rm M_\odot}

\newcommand{\mpch}{\>h^{-1}{\rm {Mpc}}}

\usepackage{amsmath,bm}
\usepackage{xcolor}

\submitjournal{ApJ}

\shorttitle{Modeling the Projected Galaxy Clustering in Photometric Surveys}
\shortauthors{Wang et al.}

\begin{document}

\title{Accurate Modeling of the Projected Galaxy Clustering in Photometric Surveys: I. Tests with Mock Catalogs}

\correspondingauthor{Xiaohu Yang, Y.P. Jing}
\email{xyang@sjtu.edu.cn, ypjing@sjtu.edu.cn}

\author[0000-0002-4073-5234]{Zhaoyu Wang}
\affiliation{Department of Astronomy, School of
  Physics and Astronomy, and Shanghai Key Laboratory for
Particle Physics and Cosmology, Shanghai Jiao Tong University,
Shanghai 200240, China}

\author[0000-0003-1132-8258]{Haojie Xu}
\affiliation{Department of Astronomy, School of
  Physics and Astronomy, and Shanghai Key Laboratory for
Particle Physics and Cosmology, Shanghai Jiao Tong University,
Shanghai 200240, China}

\author[0000-0003-3997-4606]{Xiaohu Yang}
\affiliation{Department of Astronomy, School of
  Physics and Astronomy, and Shanghai Key Laboratory for
Particle Physics and Cosmology, Shanghai Jiao Tong University, 
Shanghai 200240, China}
\affiliation{Tsung-Dao Lee Institute,
 Shanghai Jiao Tong University, Shanghai 200240, China}

\author[0000-0002-4534-3125]{Y.P. Jing}
\affiliation{Department of Astronomy, School of
  Physics and Astronomy, and Shanghai Key Laboratory for
Particle Physics and Cosmology, Shanghai Jiao Tong University,
Shanghai 200240, China}
\affiliation{Tsung-Dao Lee Institute,
 Shanghai Jiao Tong University, Shanghai 200240, China}

\author[0000-0003-4936-8247]{Hong Guo}
\affiliation{Key Laboratory for Research in Galaxies and Cosmology, 
Shanghai Astronomical Observatory, Shanghai 200030, China}

\author[0000-0003-1887-6732]{Zheng Zheng}
\affiliation{Department of Astronomy, School of
  Physics and Astronomy, and Shanghai Key Laboratory for
Particle Physics and Cosmology, Shanghai Jiao Tong University,
Shanghai 200240, China}
\affiliation{Tsung-Dao Lee Institute, Shanghai Jiao Tong University,
  Shanghai 200240, China}
\affiliation{Department of Physics and Astronomy, University of Utah,
  115 South 1400 East, Salt Lake City, UT 84112, USA}

\author[0000-0001-6966-6925]{Ying Zu}
\affiliation{Department of Astronomy, School of
  Physics and Astronomy, and Shanghai Key Laboratory for
Particle Physics and Cosmology, Shanghai Jiao Tong University,
Shanghai 200240, China}

\author[0000-0002-1324-0893]{Zhigang Li}
\affiliation{Department of Astronomy, School of
  Physics and Astronomy, and Shanghai Key Laboratory for
Particle Physics and Cosmology, Shanghai Jiao Tong University,
Shanghai 200240, China}

\author[0000-0002-4718-3428]{Chengze Liu}
\affiliation{Department of Astronomy, School of
  Physics and Astronomy, and Shanghai Key Laboratory for
Particle Physics and Cosmology, Shanghai Jiao Tong University,
Shanghai 200240, China}

\begin{abstract}
We develop a novel method to explore the galaxy-halo connection using the galaxy imaging surveys by modeling the projected two-point correlation function measured from the galaxies with reasonable photometric redshift measurements. By assuming a Gaussian form of the photometric redshift errors, we are able to simultaneously constrain the 
halo occupation distribution (HOD) models  and the effective photometric redshift uncertainties. Tests with mock galaxy catalogs demonstrate that this method can successfully recover (within $\sim 1\sigma$) 
the intrinsic large-scale galaxy bias, as well as the HOD models  
and the effective photometric redshift uncertainty. This method
also works well even for galaxy samples with 10 per cent catastrophic photometric redshift errors.
\end{abstract}


\section{Introduction}
\label{sec:intro}

The modern galaxy formation and evolution theory states that
galaxies form and evolve in the dark matter halos
\citep[e.g.][]{white_galaxy_formation, 2010gfe..book.....M}.
There are multiple ways to establish the so-called
galaxy-halo connection. The most straightforward but 
computationally expensive approach is 
the hydrodynamical simulations in a cosmological volume
\citep[e.g.][]{1996ApJS..105...19K, 2005Natur.435..629S, illustris}, 
which put in the complex baryonic physics of galaxy formation and evolution, 
such as the stellar evolution, gas heating/cooling, and active
galactic nuclei feedback. A more computationally economical method is using
the semi-analytical models
\citep[SAMs;][]{1993MNRAS.264..201K, 2005ApJ...631...21K, sam_guo,
  2017ApJ...846...66L}, which is built on the halo merger trees
extracted from $N$-body simulations \citep{2008MNRAS.383..557P,
  2014MNRAS.440..193J, 2018arXiv180900523C}. The galaxy-halo connection
can also be established in a statistical data-driven  way, 
as in the models of halo
occupation distribution \citep[HOD;][] {jing98,
  2002ApJ...575..587B, zheng05, zehavi11, guo16, 2018MNRAS.478.2019Y,
  2018ApJ...858...30G, zu15, zu16, zu17}, the conditional luminosity
function \citep[CLF;][]{yang03, 2006MNRAS.365..842C,
  2007MNRAS.376..841V, yang12, 2015ApJ...799..130R}, and the sub-halo
abundance matching \citep[SHAM;][]{2006MNRAS.371.1173V,
  abundance_matching06, 2010ApJ...717..379B, 2010MNRAS.404.1111G,
  2012MNRAS.423.3458S, 2014MNRAS.437.3228G, 2016MNRAS.460.3100C,
  guo16, 2018arXiv180403097W}. The galaxy clustering measurements, especially
the two-point correlation functions (2PCFs), are commonly employed to constrain these model parameters.

Over the last two decades, the spatial clustering of galaxies in the local universe has been extensively studied
with the advance of large-scale redshift surveys, e.g. the
2dF Galaxy Redshift Survey \citep[2dFGRS;][]{2dFGRS} and the
Sloan Digital Sky Survey \citep[SDSS;][]{SDSS}. The clustering of galaxies
is found to strongly depend on the galaxy properties, such as
luminosity, morphology, color, and spectral type
\citep[e.g.][]{jing98, 2004MNRAS.350.1153Y, 2005ApJ...633..560E,
  2005ApJ...630....1Z,zehavi11, licheng06, guo14, guo15, 2018ApJ...858...30G,
  2016ApJ...833..241S, xu16, xu18}.
The different halo models aforementioned have been successfully applied to
interpret the observed galaxy clustering measurements, as well as extract the information of the connection between galaxy properties and those of halos.

Galaxy clustering at relatively higher redshifts (e.g., $z\sim1$) has also
been studied using deep but small-area spectroscopic redshift surveys \citep[e.g.,][]{deep2_clustering1,deep2_clustering2,vipers_clustering}.
However, due to the limited sample volumes of these high-redshift
surveys, the measured galaxy clustering signals, especially on large scales, are severely
degraded by the sampling variance. 
Two methods are generally applied to infer galaxy-halo connection at these high redshifts to increase the signal-to-noise ratios of the clustering measurements. One is using the cross-correlation between the photometric and spectroscopic samples \citep[e.g.,][]{Masjedi06,2009MNRAS.399.2279M,2011ApJ...731..117H,wangwenting11} and the other is directly modeling the angular galaxy clustering measurements \citep[e.g.][]{2012A&A...542A...5C,HSC_clustering16, 2018PASJ...70S..11H, 2018ApJ...853...69C, 2018PASJ...70S..33H}.
However, both methods have their own limitations. The cross-correlation method is limited by the size of the spectroscopic sample and needs careful treatment of the interlopers, while the galaxy-halo connection in the angular clustering method is less well constrained due to the lack of redshift information. 

In order to leverage the large sky coverage and accurate photometric redshift estimation from the next-generation galaxy surveys, we develop a new method to directly model the projected 2PCFs of galaxies measured with the photometric redshifts. Under the HOD framework, by jointly modeling the projected 2PCFs with different integration depths, we are able to simultaneously constrain the HOD parameters and the photometric redshift uncertainties. We test our method against mock galaxy catalogs and find good agreement with the input model parameters. 

This paper is organized as follows. 
We describe the galaxy mock catalogs 
in section~\ref{sec:data} 
and present the methodology in section~\ref{sec:method}.
The main results are presented in section~\ref{sec:test}. 
We summarize our results and discuss the caveats in section~\ref{sec:summary}. Throughout the paper, we use $\log$ for base-10 logarithm.

\section{Data}
\label{sec:data}

In this section, we describe the $N$-body simulations and three galaxy mock catalogs used in this paper. The first two mocks, Mock I and Mock II, are based on the same $N$-body simulation. In Mock I, galaxies are randomly assigned to the positions of dark matter particles, while for Mock II, we populate the halos according to an input HOD model. We construct Mock III in another $N$-body simulation for a more realistic concern.

\subsection{\rm{Mock I} and \rm{Mock II}}
\label{subsec:mock1}

The simulation we use to create {\rm Mock I} and {\rm Mock II} contains $3072^{3}$ particles within a cubic box of 1200$^3h^{-3}\,\rm{Mpc}^3$. It was carried out using P$^{3}$M method \citep{Jing6620, cosmicgrowth} with a $\Lambda$CDM cosmology of $\Omega_{m}$ = 0.268,
$\Omega_{b}$ = 0.044, $\Omega_{\Lambda}$ = 0.732, $\sigma_{8}$ = 0.83
and h = 0.71. The mass resolution is $4.4\times10^{9}\msunh$. The
initial condition is generated at redshift $z_{i}$ = 144 following the
Zeldovich approximation, and with the transfer function from
\citet{ic}. Dark matter halos are identified by a friends-of-friends
algorithm \citep[FOF;][]{1985ApJ...292..371D} with a linking length b
= 0.2 while subhalos are identified using the
Hierarchical-Bound-Tracing algorithm \citep{HBT}.

Mock I is constructed by assigning galaxy positions from randomly selected dark matter particles in the simulation. Therefore, the galaxy bias shall be unity by design. In detail, we randomly select 3,000,000 dark matter particles from the simulation at the snapshot $z=0$. To mimic the uncertainty in photometric redshift estimation in observation,
we assign each mock galaxy a Gaussian perturbation to its $z$ axis coordinate (line-of-sight direction (LOS) under the plane-parallel approximation). We note that the LOS perturbations due to the redshift space distortion (RSD) effect (usually less than $10 \mpch$) can be ignored compared to those caused by the typical photometric redshift errors ($\sim 200 \mpch$), which we denote as photometric redshift uncertainty distortion (PRUD) to distinguish from the RSD effect. The PRUD effect in this paper is assumed to follow a Gaussian distribution, with a standard deviation of  $\sigma_{\rm photo}=200/\sqrt{2} \mpch$, corresponding to a typical $\sim 5\%$ photometric redshift uncertainty at $z \sim$ 1.5 \citep[e.g.,][]{VIPERS_data1}. The periodic boundary condition
is used here to ensure that the large scale structures at the boundary are not truncated.  We note that the simulation box size of 1200$\mpch$ is much larger than $\sigma_{\rm photo}$, therefore the periodic boundary condition can be safely applied here.

For {\rm Mock II}, we populate the halos from the same simulation \citep{Jing6620, cosmicgrowth} according to the HOD parameters\footnote{$\log M_{\rm min}=12.78\msunh$, ${\sigma}_{\log M}=0.68$, $\log M_{0}=12.71\msunh$, $\log M_{1}=13.76\msunh$, $\alpha=1.15$, see details of these parameters in section~\ref{subsec:HOD}.} for 
galaxy sample of $\rm{M_{r} < -21}$ \citep{zehavi11} from
the Sloan Digital Sky Survey Data Release 7 Main Galaxy Sample \citep[SDSS DR7;][]{Abazajian09}.
Each central galaxy is placed at
the potential minimum of the dark matter halo
while we assume the distribution of the satellite galaxies in the halos to follow the NFW profile \citep{NFW}.
The occupation number of satellite galaxies in each halo is assumed to
follow a Poisson distribution. 
In the end, we have 2,377,980 galaxies at $z$ = 0 in this mock and we apply the same PRUD effect as in \rm{Mock I}.

\subsection{\rm{Mock III}}
\label{subsec:simulation2}

To test our method with a more realistic galaxy catalog (hereafter Mock III), we improve over the previous two mocks by including a larger photometric redshift uncertainty, using a light-cone instead of a cubic box, and introducing some fraction of catastrophic redshift measurements. 

In detail, we include a larger PRUD effect and therefore
switch to a larger simulation box. We use the BigMDPL simulation that contains $3840^{3}$ particles with a box size of 2500${\mpch}$
\citep{Klypin16}. It adopts a
$\Lambda$CDM cosmology with parameters $\Omega_{m}$ = 0.307115,
$\Omega_{b}$ = 0.048206 , $\Omega_{\Lambda}$ = 0.692885 , $\sigma_{8}$ = 0.8228
and h = 0.6777. The (sub-)halos in this simulation are identified by the {\scriptsize ROCKSTAR} algorithm \citep{2013ApJ...762..109B}.

We choose a different set of HOD parameters\footnote{$\log M_{\rm min}=13.38\msunh$,
${\sigma}_{\log {M}}=0.69$, $\log M_{0}=13.35\msunh$,
$\log M_{1}=14.20\msunh$, $\alpha=1.09$} from the best-fitting model of $\rm{M_{r} < -21.5}$ galaxies of \citealt{zehavi11} in SDSS DR7. In this mock, we apply a larger photometric redshift error of $\sigma_{\rm photo}=400/\sqrt{2}{\mpch}$ and use the periodic boundary condition to construct the catalog. 

To imitate the effect of catastrophic photometric redshift measurements\footnote{We define the catastrophic redshifts are those galaxies with $|z_{\rm ph}-z_{\rm sp}|/(1+z_{\rm sp}) > 0.1$.}, we select 10\% of galaxies to have random positions in the simulation box. Then, we select galaxies from a light cone as:
\begin{enumerate}
  \item $2000{\mpch} \leq r \leq 4400{\mpch}$,
  \item $0^\circ \leq \alpha \leq 90^\circ$,
  \item $0^\circ \leq \delta \leq 90^\circ$.
\end{enumerate}
where $r$, $\alpha$ and $\delta$ are the comoving distance, the right ascension and the declination, respectively. However, we note that we have ingnored the redshift evolution effect and only use halos at the snapshot of $z=0$.

\section{Methodology}
\label{sec:method}
In this section, we provide the details of our method to simultaneously constrain the galaxy bias, the effective photometric redshift error, and the HOD parameters from modeling the projected 2PCFs in photometric surveys. We will construct our model step by step with Mocks I, II and III one after another. 

\subsection{Galaxy Bias and Photometric Redshift Error}
\label{subsec:bias}
Since we have the photometric redshift information available, we are able to measure the projected 2PCF in the photometric redshift space. As in the normal redshift-space, the projected 2PCF in the photometric redshift space can be measured by integrating the 3D 2PCF $\xi^{\rm obs}(r_{\rm p},r_{\rm \pi})$ along the LOS direction, 
\begin{equation}
w_{\rm p}^{\rm obs}(r_{\rm p}|r_{\rm \pi,max}) = 2\int_0^{\rm r_{\rm \pi, max}}\xi^{\rm obs}(r_{\rm p},r_{\rm \pi})d{r_{\rm \pi}}.
\end{equation}
where $r_{\rm \pi,max}$ is the maximum integration distance along the LOS. Different from the usual definition of the projected 2PCF, we put $r_{\rm \pi,max}$ as an additional parameter of $w_{\rm p}$, because here we use the different integration lengths to constrain the photometric redshift error, $\sigma_{\rm photo}$.

For studies investigating the galaxy clustering with spectroscopic redshifts \citep[e.g.,][]{2005ApJ...630....1Z,zehavi11,guo14,guo15,xu16,xu18}, the adopted ${r_{\rm \pi,max}}$ is usually around $40\sim 60\mpch$, which is much larger than the scale of the RSD effect ($\sim 10\mpch$). However, the shift of LOS positions caused by the uncertainties in the photometric redshift estimation are usually on the level of $\sim 150 \mpch$, corresponding to $\Delta z/(1+z) \sim 5\%$ \citep[e.g.][]{VIPERS_data1}. Therefore, a large value of $r_{\rm \pi,max}$ (e.g., $500\mpch$) is necessary to obtain a convergent estimation of the projected 2PCF. However, the measurement of $\xi^{\rm obs}(r_{p},r_{\pi})$ at such a large $r_{\rm \pi, max}$ will be dominated by the shot noise. 

Here, we propose a simple method to recover the intrinsic galaxy clustering and  the effective photometric redshift error, without integrating to an overwhelmly large ${r_{\rm \pi,max}}$. Previous studies on deep imaging surveys \citep[e.g.][]{VIPERS_data1} have suggested that the uncertainty in photometric redshift estimation usually follows a Gaussian distribution with a zero mean and a variance of $\sigma^{2}_{\rm photo}$, despite the existence of catastrophic photometric redshift estimations. We also assume a Gaussian distribution for the distortion to the LOS comoving distance due to the photometric redshift errors. As the sum of two independent Gaussian distributions also follows a Gaussian distribution, the separation of galaxy pairs in photometric redshift space 
can be modeled as a Gaussian distribution, with a variance of $\sigma_{\rm pair}^{2} = 2\sigma^{2}_{\rm photo}$.

Since the underlying galaxy bias is not dependent on scatter $\sigma_{\rm pair}$, we are able to constrain $\sigma_{\rm pair}$ from measuring $w_{\rm p}^{\rm obs}(r_{\rm p}|r_{\rm \pi,max})$ with different $r_{\rm \pi,max}$ values. In this paper, we adopt two sets of $r_{\rm \pi,max}$ as $50\mpch$ and 100$\mpch$. In principle, any two combinations of $r_{\rm \pi,max}$ can be used to achieve the constraints to $\sigma_{\rm pair}$.

Similar to the streaming model in the RSD effect, the 3D 2PCF $\xi^{\rm obs}(r_{\rm p},r^{\rm obs}_{\rm \pi})$ measured in photometric-redshift space can be modeled as a convolution of the \emph{real-space} 2PCF $\xi(r)$ with a Gaussian distribution of the distortion to the LOS separation, $\Delta r_{\rm \pi}$, as follows,
\begin{eqnarray}\label{eq:ximod}
  \xi^{\rm obs}(r_{\rm p},r_{\rm \pi})&=& 
  \int_{-\infty}^{\infty}\xi(r)f(\Delta r_{\rm \pi})d\Delta r_{\rm \pi} \\
  r&=&\sqrt{r_{\rm p}^2+(r_{\rm \pi}-\Delta r_{\rm \pi})^2}\\
  f(\Delta r_{\rm \pi})&=&\frac{1}{\sqrt{2\pi}\sigma_{\rm pair}}\exp(-\Delta r_{\rm \pi}^2/2\sigma_{\rm pair}^2)
\end{eqnarray}
We note that at high redshifts where the plane-parallel assumption works, the change to the projected distance $r_{\rm p}$ due to the photometric redshift error can be ignored. 

The large-scale galaxy bias $b$ is defined as the ratio between the galaxy 2PCF and that of the dark matter,
\begin{equation}\label{eq:bias}
  \xi(r) = b^2\xi_{\rm m}(r),
\end{equation}
where $\xi_{\rm m}(r)$ is the dark matter 2PCF, which is obtained by Fourier transforming the nonlinear matter power spectrum from CAMB\footnote{\url{http://camb.info/sources/}} \citep{CAMB}. Since the galaxy bias $b$ is degenerate with $\sigma_8$, we will only show the measurements of $b\sigma_8$ in the following sections. 

\subsection{HOD Model}
\label{subsec:HOD}
In order to model the projected 2PCF $w_{\rm p}^{\rm obs}(r_{\rm p}|r_{\rm \pi,max})$, we follow the HOD model framework laid out in \cite{2013MNRAS.430..725V}. We refer the readers there for model details. In brief, we separate the contribution of the average occupation number $\langle N(M_{\rm h})\rangle$ of galaxies in halos of mass $M_{\rm h}$ into those from the central and satellite galaxies \citep*{zheng07},
\begin{eqnarray}\label{eq:HOD1}
  N(M_{\rm h}) &=& N_{\rm c}(M_{\rm h}) + N_{\rm s}(M_{\rm h}),\\
  N_{\rm c}(M_{\rm h}) &=& {\frac{1}{2}}\left[1+{\rm erf}\left(\frac{\log M_{\rm h}-\log
    M_{\rm min}}{{\sigma}_{\log M}}\right)\right],\\
  N_{\rm s}(M_{\rm h}) &=& N_{\rm c}(M_{\rm h})\left(\frac{M_{\rm h}-M_{0}}{M_{1}}\right)^{\alpha}.
\end{eqnarray}
As in the traditional HOD model, the five free parameters are the central galaxy cutoff halo mass $M_{\rm min}$, width of the cutoff profile ${\sigma}_{\log M}$, satellite galaxy cutoff halo mass $M_0$, normalization $M_1$ and the high-mass end slope $\alpha$. With the additional model parameter $\sigma_{\rm pair}$, we are able to predict the observed projected 2PCF $w_{\rm p}^{\rm obs}(r_{\rm p}|r_{\rm \pi,max})$ at different $r_{\rm \pi,max}$ values. 

\subsection{A More Realistic Mock Galaxy Catalog}
\label{subsec:amrm}

In Mock III, we have assumed 10\% of galaxies to have catastrophic redshifts, therefore the measured 2PCF $\xi^\prime(r)$ would deviate from the intrinsic 2PCF $\xi(r)$. Since we use the random dark matter particles to represent the 10\% galaxies with catastrophic redshifts, $\xi^\prime$ can be estimated as \citep{1998ApJ...494L..41S},
\begin{eqnarray}
\label{real_corr}
 \xi^\prime(r) &=&
  \rm{\frac{(D^\prime-R)(D^\prime-R)}{RR}}\nonumber \\
  &=&\rm{\frac{[{\it f}D+(1-{\it f})R-R][{\it f}D+(1-{\it f})R-R]}{RR}}\nonumber\\
  &=&f^2\rm{\frac{(D-R)^2}{RR}}\nonumber \\
  &=&f^2\xi(r)
\end{eqnarray}
where $f$ is the fraction of galaxies with reasonable redshift measurements, i.e. 90\% in Mock III. `D' represents a data point and `R' denotes a random galaxy point. We will use the above equation to  model the observed 2PCF in Mock III.

\begin{figure}
\includegraphics[width =\columnwidth]{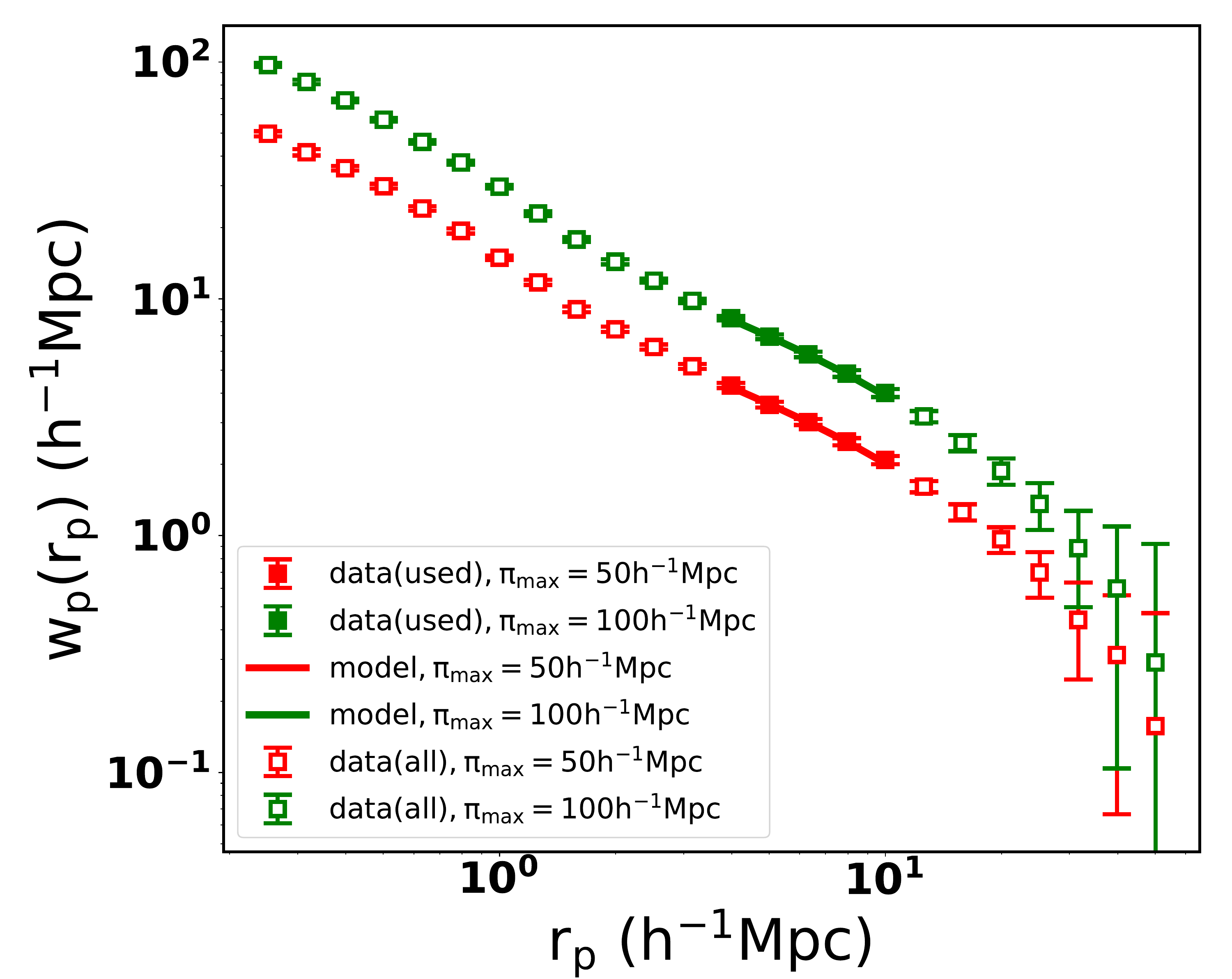}
\caption{Comparisons of measured and model  projected 2PCFs at different $r_{\rm \pi,max}$ for Mock I.  
Red (green) filled squares with errorbars are the 
measurements for $r_{\rm \pi,max} = 50 (100)\mpch$, 
and the model predictions are shown as solid lines. The open squares
are measurements for all scale, shown as a reference.}
\label{fig:wp1}
\end{figure}

\begin{figure}
\includegraphics[width=\columnwidth]{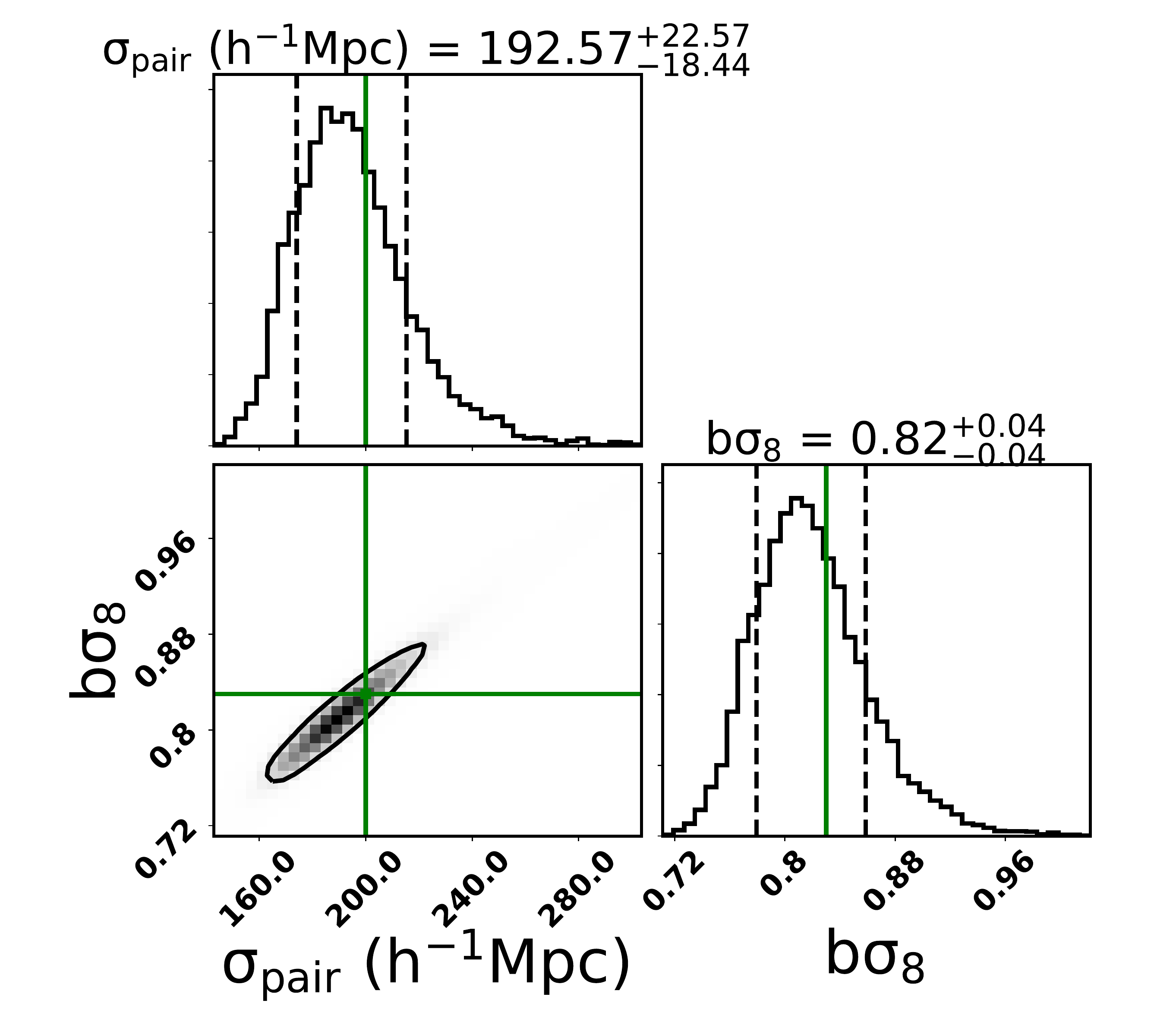}
\caption{The 2D contours and 1D distribution of the parameters $\sigma_{\rm pair}$ and $b\sigma_8$ for Mock {\rm I}. The black solid line in the contour corresponds to the 68.3\% confidence region. The vertical black dashed lines represent the 68.3\% confidence ranges for both parameters. The median and 68.3\% confidence ranges are also labeled on top of the figure. The green solid lines are the input model parameters.} 
\label{fig:fit1}
\end{figure}

\section{Results}
\label{sec:test}

In this section, we show the testing results of our method on Mocks I, II, and III.
To measure the projected 2PCF $w_{\rm p}(r_{\rm p}| r_{\rm \pi,max})$, we adopt 24 logarithmic $r_{\rm p}$bins in $ -0.6 < \log (r_{\rm p}/\mpch) < 1.7$ and linear bins in $r_{\pi}$ with $\Delta r_\pi=2\mpch$ from $0$ to $r_{\rm \pi,max} = 50$ and $100\mpch$. The error covariance matrix for $w_{\rm p}$ is estimated from using the jackknife resampling method with 100 subsamples \citep[see e.g.,][]{zehavi11,guo13,xu16,xu18}. We note that the error covariance between the projected 2PCF measurements with different
$r_{\rm \pi,max}$ values have been taken into account.

We adopt a Markov Chain Monte Carlo (MCMC) method to explore the parameter space using the $\chi^2$ that is defined as,
\begin{eqnarray}
  \chi^{2} &=& (\bm{w_{\rm p}^{\rm obs}}-\bm{w^{\rm mod}_{\rm p}})^\mathbf{T}
  {\bm C^{-1}}(\bm{w_{\rm p}^{\rm obs}}-\bm{w^{\rm mod}_{\rm p}})\nonumber\\
  &&+(n_{\rm g}^{\rm obs}-n_{\rm g}^{\rm mod})^2/\sigma_{n_{\rm g}}^2\,,
\end{eqnarray}
where ${\bm C}$ is the error covariance matrix of the data vector. $\bm{w_{\rm p}}$ includes the measurements with two sets of different $r_{\rm \pi,max}$ values, i.e., $\bm{w_{\rm p}}$=[$\bm{w_{\rm p}}(\bm{r_{\rm p}}| r_{\rm \pi,max1})$,
$\bm{w_{\rm p}}(\bm{r_{\rm p}}| r_{\rm \pi,max2})$]. The scatter $\sigma_{n_{\rm g}}$ of the galaxy sample number density $n_{\rm g}$ is also estimated with the jackknife resampling method. The quantity with subscripts of ``obs'' and ``mod'' are for the observed and model measurements, respectively. Here we place flat priors on the input parameters, with broad parameter ranges. For example, the priors on $b\sigma_8$ and $\sigma_{\rm pair}$ in Mock I are $[0,8.3]$ and $[0,1000]$, respectively. Since in Mock I, we only have two free parameters, we run 30 MCMC chains, each with 20,000 steps, while for Mock II and III we run 50,000 steps due to their more free parameters. Most of the chains converge within the first 20\% of steps. These first steps are regarded as burn-in stage and removed from our chains. In order to suppress the correlation between neighboring models in each chain, we thin the chain by a factor 10. This results in  final MCMCs consisting of about 48,000 and 120,000 independent models that sample the posterior distributions for Mock I and II (III), respectively. 

\begin{figure}
\includegraphics[width = \columnwidth]{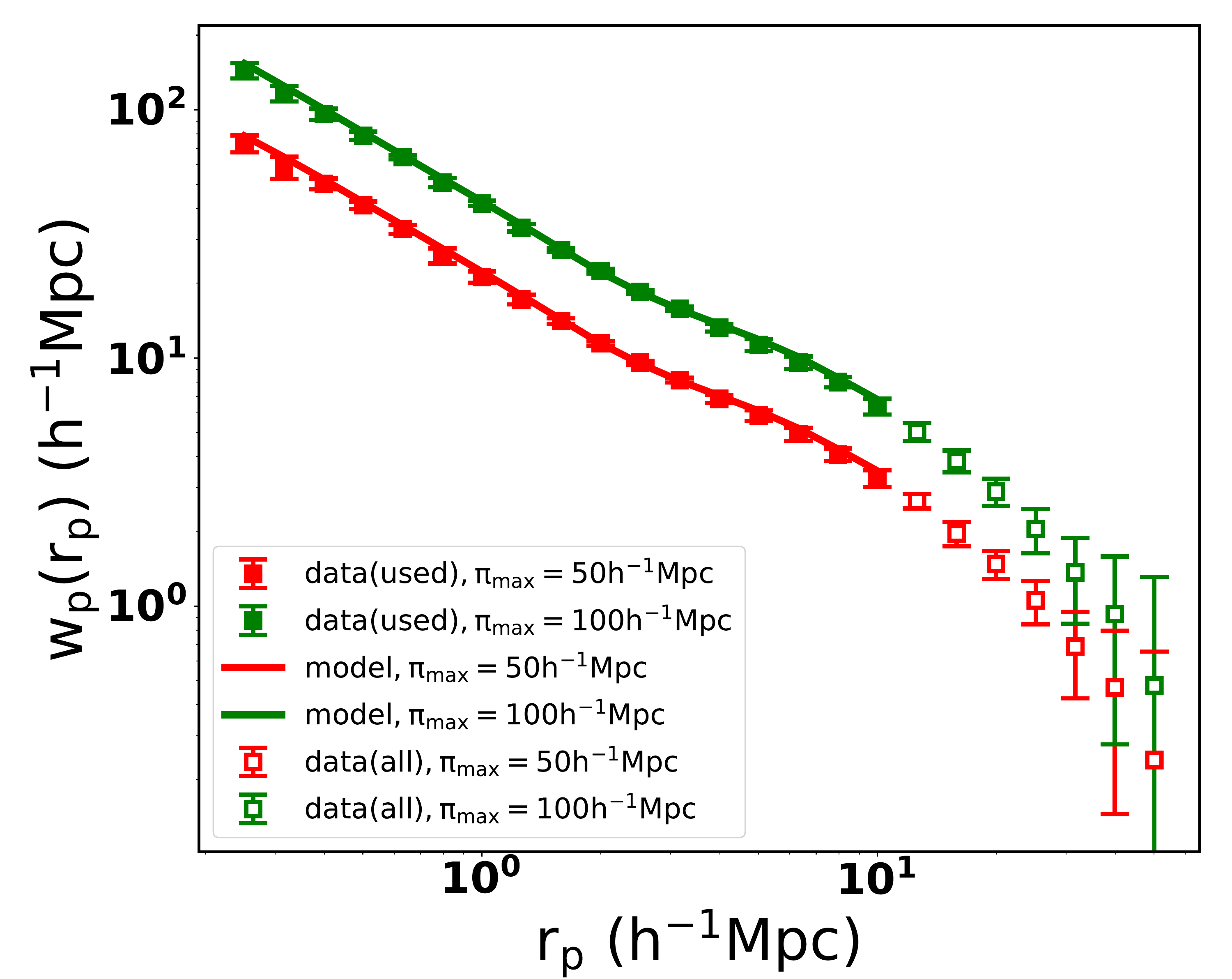}
\caption{Similar to Figure~\ref{fig:wp1}, but for Mock II.} 
\label{fig:wp2}
\end{figure}

\begin{figure*}
\includegraphics[width=\textwidth]{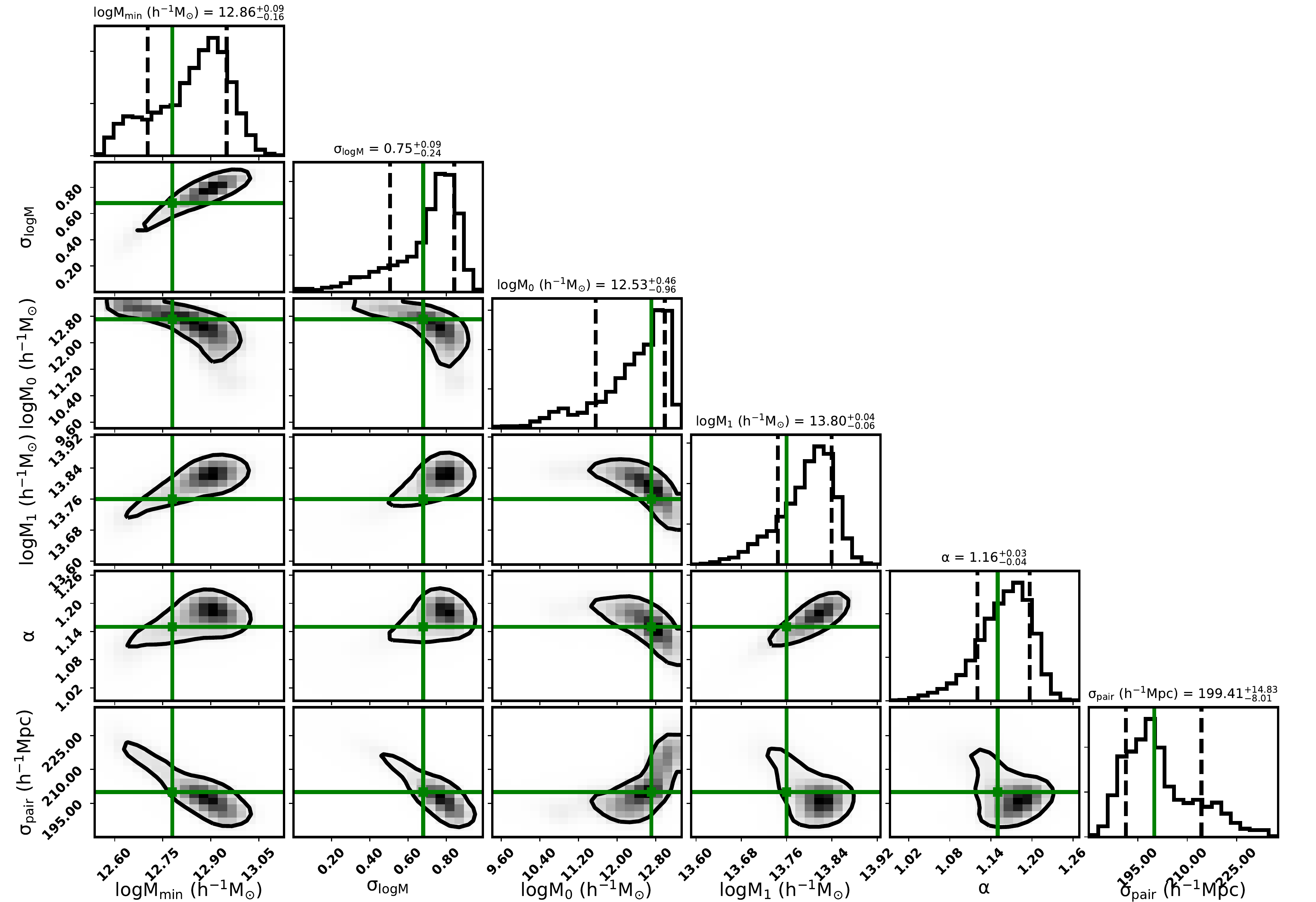}
\caption{Simiar to Figure~\ref{fig:fit1}, but for the joint probability distributions of the HOD parameters and $\sigma_{\rm pair}$ for Mock II. The best-fitting model parameters with the 1$\sigma$ ranges are also labels.}
\label{fig:fit2}
\end{figure*}

\subsection{Mock I: Large-scale Galaxy Bias} \label{subsec:cac}

In Mock I, we randomly select the dark matter particles to represent galaxies in order to check whether we can recover the correct galaxy bias $b$ and photometric redshift error $\sigma_{\rm pair}$. Therefore, we only use the $w_{\rm p}$ measurements at relatively large scales of $4< r_{\rm p}< 10\mpch$, shown as the filled squares in Figure~\ref{fig:wp1}).
We exclude the measurements on larger scales
due to their low signal-to-noise ratios.

It is clearly shown in Figure~\ref{fig:wp1} that $w_{\rm p}$ measurements with $r_{\rm \pi,max}=100\mpch$ have much higher clustering amplitudes compared to those with $r_{\rm \pi,max}=50\mpch$, implying that the measurements of $w_{\rm p}$ are not converged with relatively small values of $r_{\rm \pi,max}$. Especially in the situation of the large photometric redshift errors, galaxy pairs with large LOS distances still have a significant contribution to the clustering measurements. It emphasizes the importance of integrating to a extremely large $r_{\rm \pi,max}$ for traditional $w_{\rm p}$ measurements. While in our method, we can use such differences in the clustering measurements to estimate $\sigma_{\rm pair}$.

For this simplest mock, we have only two free parameters of $b\sigma_8$ and $\sigma_{\rm pair}$ as in Eqs.~(\ref{eq:ximod}) and~(\ref{eq:bias}). We show the best-fitting model predictions of $w_{\rm p}(r_{\rm p})$ as the solid lines in Fig.~\ref{fig:wp1}. The joint probability distribution of $\sigma_{pair}$ and $b\sigma_8$ is shown in Fig.~\ref{fig:fit1}. The input model parameters (green horizontal and vertical lines) are very well recovered within the 1$\sigma$ parameter distributions, demonstrating the validity of our method.

\subsection{Mock II: HOD Model Parameters}
\label{subsec:hmpc}

In Mock {\rm II}, as we have assigned galaxies to halos using the HOD model, we have five HOD model parameters and an additional one of $\sigma_{\rm pair}$. To include the small-scale information in this mock to constrain the HOD model, we use the $w_{\rm p}$ measurements with $0.25\mpch<r_{\rm p}<10\mpch$, i.e. we have 17 data points of $w_{\rm p}$ for each $r_{\rm \pi,max}$. The best-fitting model and joint probability distributions of the model parameters are shown in Figures~\ref{fig:wp2} and~\ref{fig:fit2}, respectively.

All of the HOD model parameters are reasonably recovered within the 1$\sigma$ probability distributions. We also note that the recovered $\sigma_{\rm pair}$ is in remarkable agreement with the input parameter, making our method a promising way of constraining the HOD, as well as testing the accuracy of photometric redshifts. 

\begin{figure}
\includegraphics[width=\columnwidth]{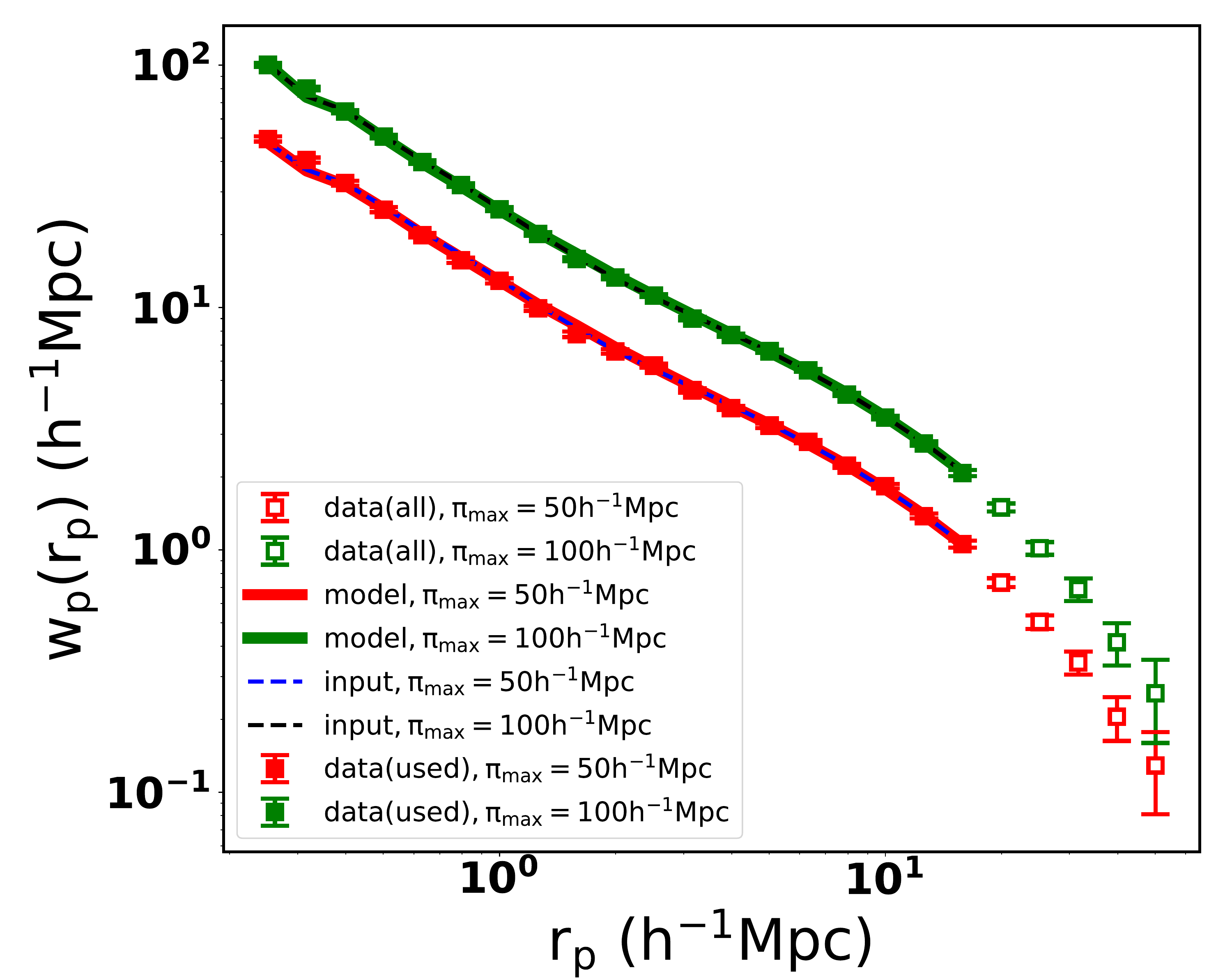}
\caption{Similar to Figure~\ref{fig:wp2}, but for Mock III. For comparison, we also show the HOD model predictions with the input parameters as the dotted lines of different colors.}
\label{fig:wp3}
\end{figure}
\begin{figure*}
\includegraphics[width=\textwidth]{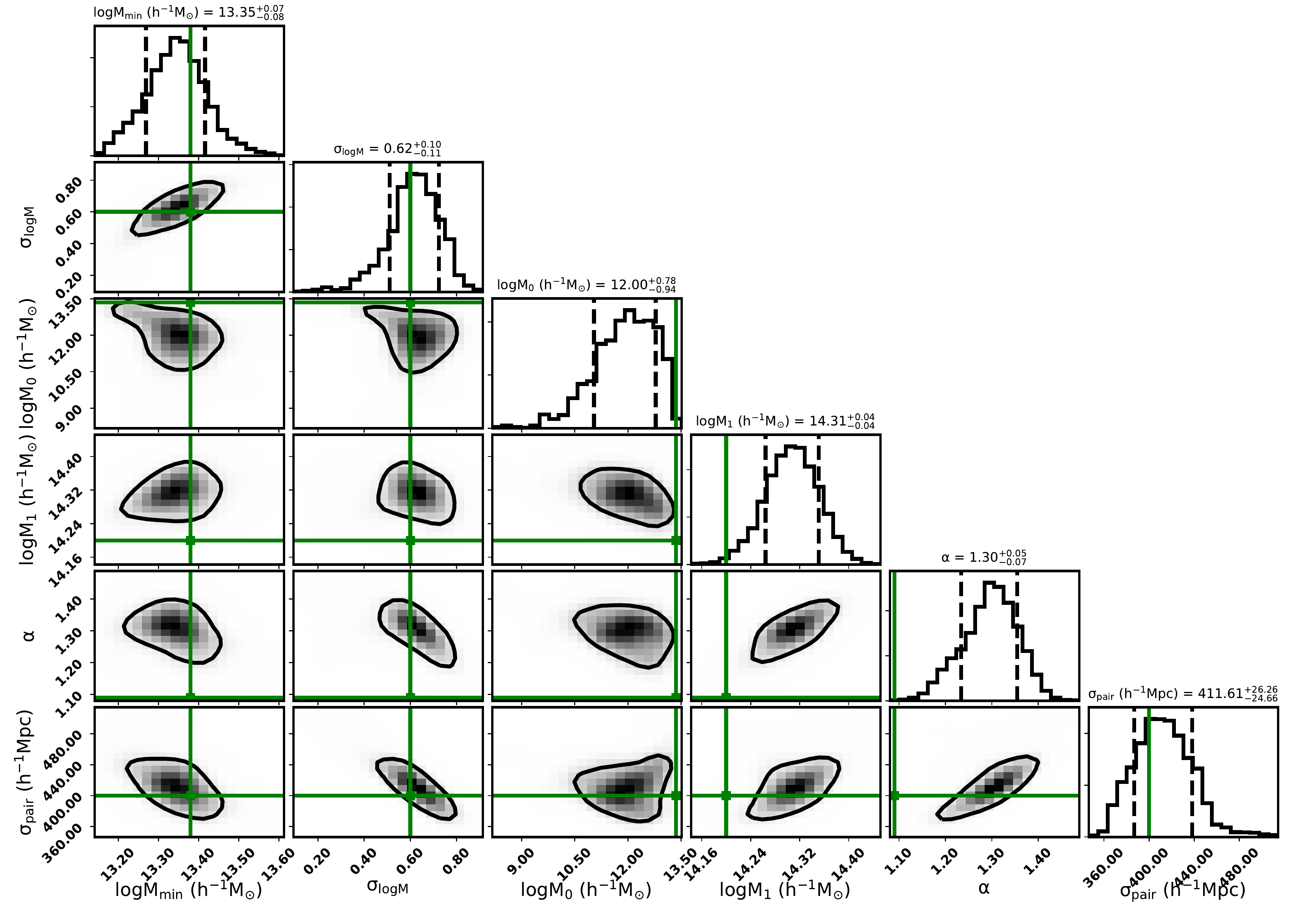}
\caption{
Similar to Fig.~\ref{fig:fit2}, but for Mock III.} 
\label{fig:fit3}
\end{figure*}

\subsection{Mock III: More Realistic Case}

Now we turn to the much more realistic Mock III. The best-fitting model and joint probability distributions are shown in Figures~\ref{fig:wp3} and~\ref{fig:fit3}, respectively. The two model parameters for the central galaxy occupation number, $M_{\rm min}$ and $\sigma_{\log M}$, are well recovered as in the case of Mock II. However, the best-fitting model parameters for satellite galaxies are not well recovered. It overpredicts the satellite occupation parameters of $M_0$ and $M_1$ and underpredicts that of $\alpha$. 

To figure out the origin of such a discrepancy, we compare the best-fitting $w_{\rm p}(r_{\rm p})$ to the model predictions using the input HOD parameters (over-plotted as the thin lines in Fig ~\ref{fig:wp3}), and find that they are almost indistinguishable with each other. Since Mock III adopts the HOD parameters from \cite{zehavi11} for the very luminous galaxies of $M_{r}<-21.5$, where the satellite galaxy fraction is only 9\%, the strong degeneracy between the satellite HOD parameters make it difficult to exactly recover the input model parameters. Meanwhile, the central galaxy occupation parameters can still be well constrained from the clustering measurements, as well as the photometric errors.

\begin{figure}
\includegraphics[width=\columnwidth]{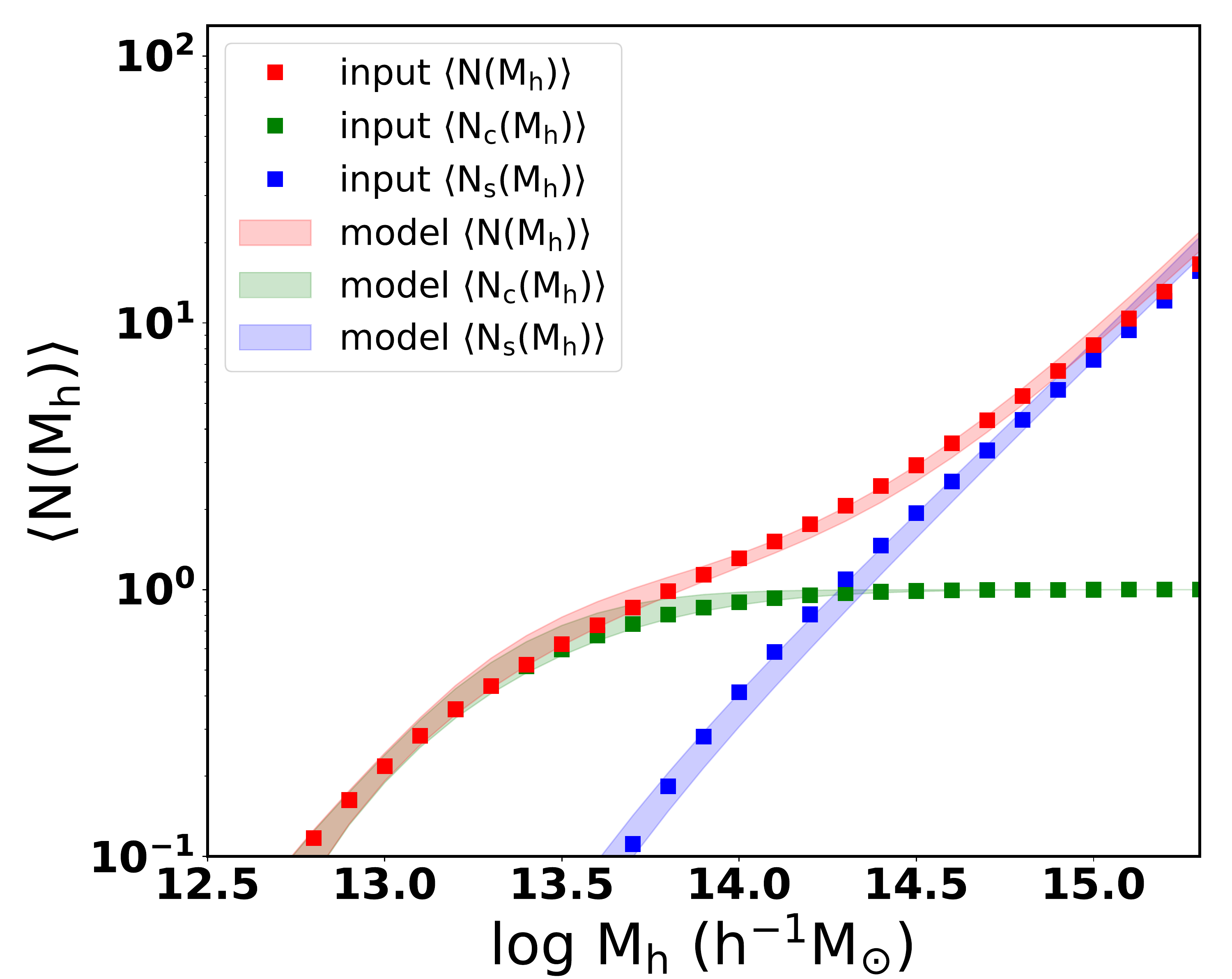}
\caption{Comparison between input (filled squares) halo occupation functions and those of the 1$\sigma$ distributions of the model predictions (shaded regions). We show the occupation functions for all (red), central (green) and satellite (blue) galaxies, respectively.} 
\label{fig:occupation}
\end{figure}

Rather than the HOD parameters, we show in Figure~\ref{fig:occupation} the direct comparisons between the observed and model {\it occupation functions} for both central and satellite galaxies. 
Here we do see that both the central and satellite galaxy occupation functions are well recovered with respect to their input ones. It is therefore important to emphasize that in the HOD model constraints using clustering measurements, it would be safer to compare the HOD functions rather than individual parameters. 

In summary, with the above three mock tests, we demonstrate that by measuring $w_{\rm p}(r_{\rm p})$ at different $r_{\rm \pi,max}$ values in the photometric redshift surveys, we are able to reasonably constrain the HOD models as well as provide independent constraint on the overall photometric redshift error. 

\section{Conclusions and Discussions}
\label{sec:summary}

In this paper, we develop a method to simultaneously constrain the large-scale galaxy bias, the HOD models, and photometric redshift uncertainty by joint modeling the projected 2PCFs with multiple integration depths along the LOS direction. With the assumption that the difference between the photometric and true redshifts of galaxies follow a Gaussian distribution, we are able to model the observed projected 2PCFs in photometric redshift surveys under the HOD model framework. 

We have tested our method in three mock galaxy catalogs at different complexity levels: (1) Mock I is constructed by randomly selecting 
the dark matter particles to represent galaxies; (2) In Mock II, we assign galaxies to halos in the simulation box using a set of HOD model parameters; (3) In Mock III, we select galaxies in a fixed light cone, introduce larger photometric redshift errors, and even include 10\% catastrophic redshift measurement errors. The best-fitting models in all three mocks agree reasonably well with the input model parameters. The recovered photometric redshift errors are in very good agreement with the expected values, making our method an independent test of the accuracy of the photometric redshifts in the survey pipeline output. It is especially promising that our method can recover the HOD model parameters with a reasonable accuracy from using only the photometric galaxy samples.

\cite{2009MNRAS.399.2279M} proposed a method to infer the projected two-point cross-correlation function between a photometric sample and a spectoscopic sample by using the probability distribution function of the photometric redshifts, which significantly enhances the clustering signal.  \cite{2011ApJ...731..117H} further applied this method to the quasar clustering in the Bo\"otes multi-wavelength survey. Such a method is useful when large spectroscopic samples are available at the redshifts of interest. However, high-redshift galaxy samples with spectroscopic redshifts are usually limited in sample size and area. Therefore, our method focuses on the auto-correlation function within the photometric samples and we adopt a forward modeling way to account for the photometric redshift uncertainties in the halo model rather than recovering the intrinsic real-space correlation function from observation.

\cite{2018ApJ...853...69C} proposed a similar Gaussian error kernal in their modeling of angular galaxy clustering measurements incorporated with photometric redshift errors. Compared to the traditional halo modeling of the galaxy angular clustering measurements, the main advantage of modeling the projected 2PCF  in the photometric redshift surveys is that the redshift extent in angular clustering measurements is usually much larger than the photometric redshift errors, which leads to a large projection effect, reducing the number of effective modes and thus weakening the constraints on the galaxy-halo connection. Moreover, modeling the projected 2PCF is much simpler than the angular one, where the conversion from the real-space 2PCF to the angular one is necessary \citep[see e.g.,][]{2012A&A...542A...5C,2018ApJ...853...69C}. Due to the strong correlation between $\sigma_{\rm pair}$ and $b\sigma_8$, the photometric redshift error can be better constrained in our modeling of the projected 2PCF. 

At the moment, there are many finished and on-going galaxy photometric surveys with various depths, e.g., the Canada-France-Hawaii Telescope Legacy Survey \citep[CFHTLS;][]{2012SPIE.8448E..0MC}, the Dark Energy Survey \citep[DES;][]{DES}, the Kilo-Degree Survey \citep[KiDS;][]{KIDS}, the The Dark Energy Camera Legacy Survey \citep[DECaLS;][]{DESI_image}, the Hyper Suprime-Cam \citep[HSC;][]{HSC}, as well as the next-generation surveys, such as the Large Synoptic Survey Telescope \citep[LSST;][]{2018AJ....155....1G}. Our method is potentially very useful for understanding the galaxy-halo connection at different cosmic epochs with these large-scale galaxy photometric surveys.

However, we also note that we have assumed a simple Gaussian form for the photometric redshift error. It is possibly different from the real distribution of the photometric redshift errors in the surveys. For example, \cite{guo15} found that the redshift error in SDSS in fact follows a Gaussian-convolved Laplace distribution. But the Gaussian distribution can still be used as a first-order approximation as shown in \cite{VIPERS_data1} and \cite{2018ApJ...853...69C}. Our model can be further improved in future by taking into account the fact that the photometric redshift error may be dependent on the galaxy luminosity and redshift. By incorporating the simulation-based halo modeling method as proposed in \citet{2016MNRAS.458.4015Z}, we will be able to provide more accurate constraints to the HOD parameters, which will be explored in a subsequent paper of applying our method to the real photometric surveys (Xu et al. in preparation).

\acknowledgments
We thank the anonymous referee for the helpful comments that significantly improve the presentation of this paper.
ZYW thanks Pengjie Zhang and Kai Wang for useful discussions and help.
This work is supported by the 973 Program (Nos. 2015CB857002,
2015CB857003), national science foundation of China (grant
Nos. 11833005, 11890691, 11890692, 11533006, 11621303) and Shanghai Natural
Science Foundation, Grant No. 15ZR1446700. We also thank the support
of the Key Laboratory for Particle Physics, Astrophysics and
Cosmology, Ministry of Education.

The CosmoSim database used in this paper is a service by the Leibniz-Institute for Astrophysics Potsdam (AIP).
The MultiDark database was developed in cooperation with the Spanish MultiDark Consolider Project CSD2009-00064.

\bibliography{photo}

\end{document}